\documentclass[twocolumn,showpacs,preprintnumbers,amsmath,amssymb]{revtex4}

\usepackage{graphicx}
\usepackage{dcolumn}
\usepackage{bm}

\begin{document}
\preprint{APS/123-QED}
\title{Continuous melting of a driven two-dimensional flux lattice with strong pins}
\author{L. Fruchter}%
\affiliation{Laboratoire de Physique des Solides, C.N.R.S.
Université Paris-Sud, 91405 Orsay cedex, France }
\date{\today}
%
\begin{abstract}
The phase diagram of a driven two-dimensional vortex
lattice in the presence of dense quasi-point pins is
investigated. The transition from the crystal to the liquid is
found continuous at intermediate inductions. The correlations in
the pseudo random force that allow for an uncomplete unbinding of
the dislocations is proposed as a key mechanism to account for the
continuous transition.
\end{abstract}
%
%

\maketitle

\section{Introduction}
It has long been noticed that a driven elastic lattice driven at
zero temperature may experience pinning as an effective shaking
temperature, due to randomly induced displacements of the lattice
nodes. This powerful analogy allows for the prediction of some
properties of the driven lattice from the common phase diagram of
particles with a repulsive interaction. In particular, a dynamic
melting transition is predicted and observed
numerically\cite{koshelev94}. However, strongly disordered
systems, as obtained in the presence of strong quasi-point pins,
may move aside from this picture. Indeed, the thermal analogy may
break down for the moving crystal due to temporal correlations of
the pseudo thermal Langevin force, a situation which is
encountered in the case of heterogeneous pinning involving
plastic flow channels\cite{balents98}. As a consequence, while
this analogy accounts for the existence of a first order-like
dynamic transition, the driven phases may differ from their
thermodynamic analogues. Several examples of such exotic phases
have been given in refs. \cite{moon96,ryu96,olson98,fangohr01}.
The anisotropy of the pinning potential, once tilted by the
external force, is essential to the formation of these
phases\cite{balents98,ledoussal98}. Here, the dynamic transition
is examined in detail for a simple system of dense quasi-point
pins with positional disorder. For part of the phase diagram, it
is found that there is a continuous transition between the
crystal and the liquid, through what might be called a liquid
crystal, whereas the transition is first order like for the rest
of the phase diagram. In the case of the continuous transition,
disclinations tend to form chains, which likely arise from the
correlations of the pseudo random force which are specific to the
driven lattice.

\section{Experiment}
The numerical sample used here is the one in ref.~
\cite{fruchter02}. It is the one of two dimensional particles
with a repulsive interaction, interacting also with a random
attractive potential. The sample mimics a two dimensional vortex
lattice, or a three dimensional rigid vortex lattice, in the
presence of strong pins, as can be created by heavy ions
irradiation. Adopting the terminology of superconductors, the
vortex density is set by the magnetic induction, $B$, as $n =
a_0^{-2} = B / \Phi_0 $ - $\Phi_0$ being the flux quantum carried
by each vortex ($2\;10^{-7}\;G cm^2$). The repulsive force between
vortices is taken as :

\begin{equation}
f_{vv}(r)=\left(A_{V}/\lambda\right) K_{1}\left(r/\lambda\right)
\label{fvv}
\end{equation}

where $K_{1}$ is a Bessel function, behaving as $\ln\;r^{-1}$ at
short distance and $r^{-1/2}\exp(-r)$ at large distance. To keep
computation tractable, the repulsive force is cut smoothly at a
distance $11 a_0$, which insures that each particle interacts with
many of its closest neighbors.

The short range potential originates from strong pins randomly
distributed in the sample, each creating the attractive force in
the range $r_{P}$:

\begin{equation}
f_{p}(r) = \left(2\;A_{P}/r_{P}\right)\:(r/r_{P}),\: \textrm{for
$r \leq r_{P}$};\;
 0  \;\textrm{for $r > r_{P}$}
\label{fp}
\end{equation}
All pins are identical and the randomness of the potential
originates from the pins position only.

In the rest, driving current densities are normalized to the
single vortex critical current density, $J_c =
2\;A_{P}/r_{P}\;\Phi_0$. The density of the pinning sites,
relative to that of the vortices, $B_\Phi/B$, is constant and
equal to $12$. The pinning potential range, relative to the
vortex average separation, is also constant and equal to $r_P/a_0
= 5.5 \;10^{-2}$, as well as the reduced force magnitude,
$\lambda A_{P}/r_{P} A_{V} = 20$. As a consequence, using $a_0$ as
the length scale, the different numerical experiments made for
different values of the induction $B$ only differ by the reduced
vortex interaction length, $\lambda/a_0$, where it was set
$\lambda=1400\;\AA$. As in \cite{fruchter02}, the total force on
each vortex, originating from its neighbors, a possible pinning
site at the vortex location and the uniform external force is
computed at each time step. Vortices which are not pinned are then
moved on a time interval small enough so that their motion is
small compared to all characteristic lengths. The boundary
conditions are periodic along the driving force direction. A
large area free from any pinning site is kept at the sample edges
parallel to the vortex motion, where a perfect hexagonal lattice
is obtained under the action of the external magnetic
pressure\cite{fruchter02}. In this way, the measurements actually
sample the driven phase embbeded in the crystal. Whereas such an
interface may promote the formation of the ordered phase in the
case of a first order transition and for finite samples, in the
case of a continuous transition, as will be discussed later, the
interface probably induces an interfacial layer only. In the
following, samples far enough from the interface are considered
and their uniformity is an indication that finite size effects
are not playing a major role when a continuous transition is
observed.

Experiments are carried out for different values of the induction
and of the external force. After a stationary state is obtained
(characterized by a steady average velocity), a snapshot of the
moving lattice is recorded, on which a Delaunay triangulation is
performed. Positive and negative disclinations (vortices with
coordinance 5 and 7), either free or forming dislocations by
pairs\cite{nelson79} are counted. Samples typically enclose
$7000$ vortices and $4\;10^4$ pins.

\label{results}

\section{results and discussion}

\begin{figure}

\resizebox{0.5\textwidth}{!}{%
  \includegraphics{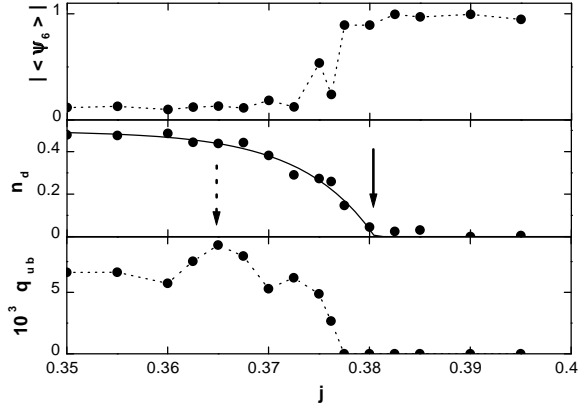}
} \caption{$B = 3950 \; Oe$. From top to bottom : Hexatic
parameter, concentration of defects (sites with coordination
number not equal to 6), concentration of free disclinations
(defects bound to sites with coordination number 6 only). The
line is the fit described in the text; the full line arrow
indicates the onset for the defects creation, as obtained from
this fit; the dotted one is the melting point as obtained in
Fig.\ref{length}.} \label{159}
\end{figure}

\begin{figure}
\resizebox{0.5\textwidth}{!}{%
  \includegraphics{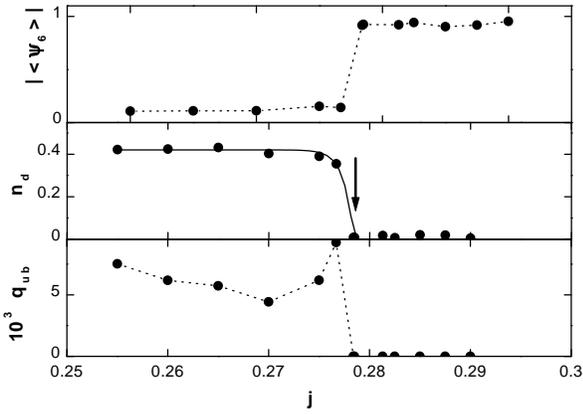}
} \caption{$B = 10^{4} \; Oe$.} \label{155}
\end{figure}

As shown in ref.\cite{fruchter02}, as the driving force
decreases, the system evolves from a moving crystal to an
amorphous phase. Contrasting with the results in
\cite{moon96,ryu96,olson98}, the high velocity phase does not
show here smectic ordering, as evidenced from the diffraction
pattern : this comes from the small ratio $r_p/a_0$ and from the
fact that the tilted pinning potential shows here a moderate
anisotropy on the scale of $a_0$. There is no attractive
interaction between the vortices, which would allow for a
transition between a liquid and a gas. However, considering the
comparable densities of the crystal and the less ordered phase,
as well as the strong interactions between the vortices in the
amorphous phase, it must obviously be called a 'liquid phase'. As
evidenced in Fig.\ref{159} and \ref{155} and the inspection of
the average hexatic parameter,$|<\Psi_{6}>|=
|<1/c_\alpha\;\sum_{\beta=1,c_\alpha}
e^{6\;i\;\theta_{\alpha,\beta}}>_{\alpha}|$ where is $c_\alpha$
the coordination number for vortex $\alpha$ and
$\theta_{\alpha,\beta}$ is the angle of the bond between
neighboring vortices $\alpha$ and $\beta$, some residual
orientational correlation is retained for low $j$ ($|<\Psi_{6}>|
\simeq 0.1$), which justifies to call the low $j$ phase an
'hexatic liquid'\cite{ryu96}.

\begin{figure}
\resizebox{0.5\textwidth}{!}{%
  \includegraphics{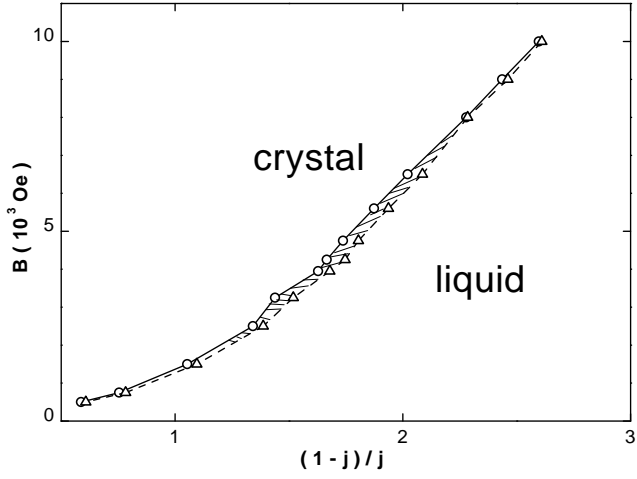}
} \caption{Dynamic phase diagram of the driven lattice. Circles:
$j = j_o$, triangles : $j = j_o-\delta$.} \label{phasediag}
\end{figure}

I now examine in more detail the transition between the crystal
and the liquid. For all systems, the concentration of defects
exhibits a clear onset upon decreasing the driving force, similar
to the one reported in \cite{koshelev94}. However, depending on
the magnetic pressure, a discontinuous or gradual rise of this
concentration is observed. This may be seen in Figs. \ref{159} and
\ref{155} obtained for two different magnetic inductions, which
clearly exhibit respectively a gradual and a step increase of the
number of defects. In order to quantify this observation, the
defects concentration was fitted with an exponential, $n_d
\propto 1 - \exp [(j_o -j)/\delta]\; (j < j_o) $, yielding the
onset, $j_o$, and a width for the transition to the liquid phase,
$\delta$. A phase diagram similar to the temperature-density
representation for the thermodynamics may be obtained, using the
theory for the equivalent 'shaking temperature'\cite{koshelev94}.
It should be stressed that this representation is qualitative
only, considering the reservations made in
ref.\cite{koshelev94,balents98} (mainly, the perturbative approach
from the uniform velocity which rules out plasticity, and the
observation that the effective temperature differs for the
fluidlike motion and the coherent one). Also, the equivalent
temperature in \cite{koshelev94}, must be modified to account for
the proximity of the flux flow to the flux creep crossover.
Imposing for the equivalent temperature to be proportional to the
potential well depth when $j_o \rightarrow 1$ and $B \rightarrow
0$, $T \propto (1-j)$, and using $T \propto j^{-1}$ from
\cite{koshelev94}, a phase diagram is obtained in the $B$ vs
$(1-j)/j$ representation. The onset for the defects concentration
($j_o$), as well as the location where it saturates
($j_o-\delta$), are plotted in this way in Fig.\ref{phasediag}.
Clearly, there is a range of magnetic induction for which a
regime, intermediate between the moving crystal and the hexatic
liquid, can be found. The existence of such a regime was already
pointed out in \cite{fruchter02}.

\begin{figure}
\resizebox{0.5\textwidth}{!}{%
  \includegraphics{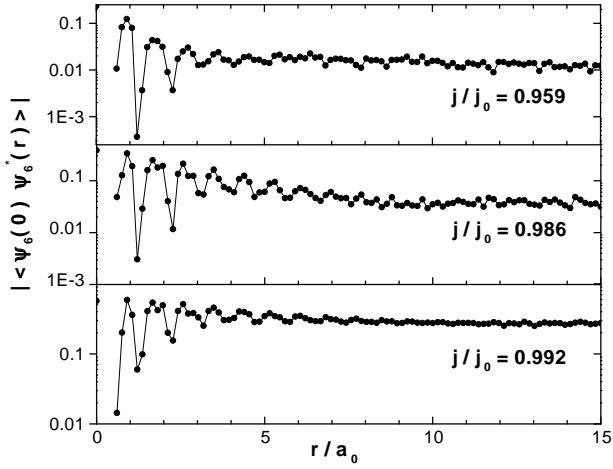}
} \caption{The correlation function for the hexatic parameter.
($B = 3950$ Oe)} \label{psi6}
\end{figure}

\begin{figure}
\resizebox{0.5\textwidth}{!}{%
  \includegraphics{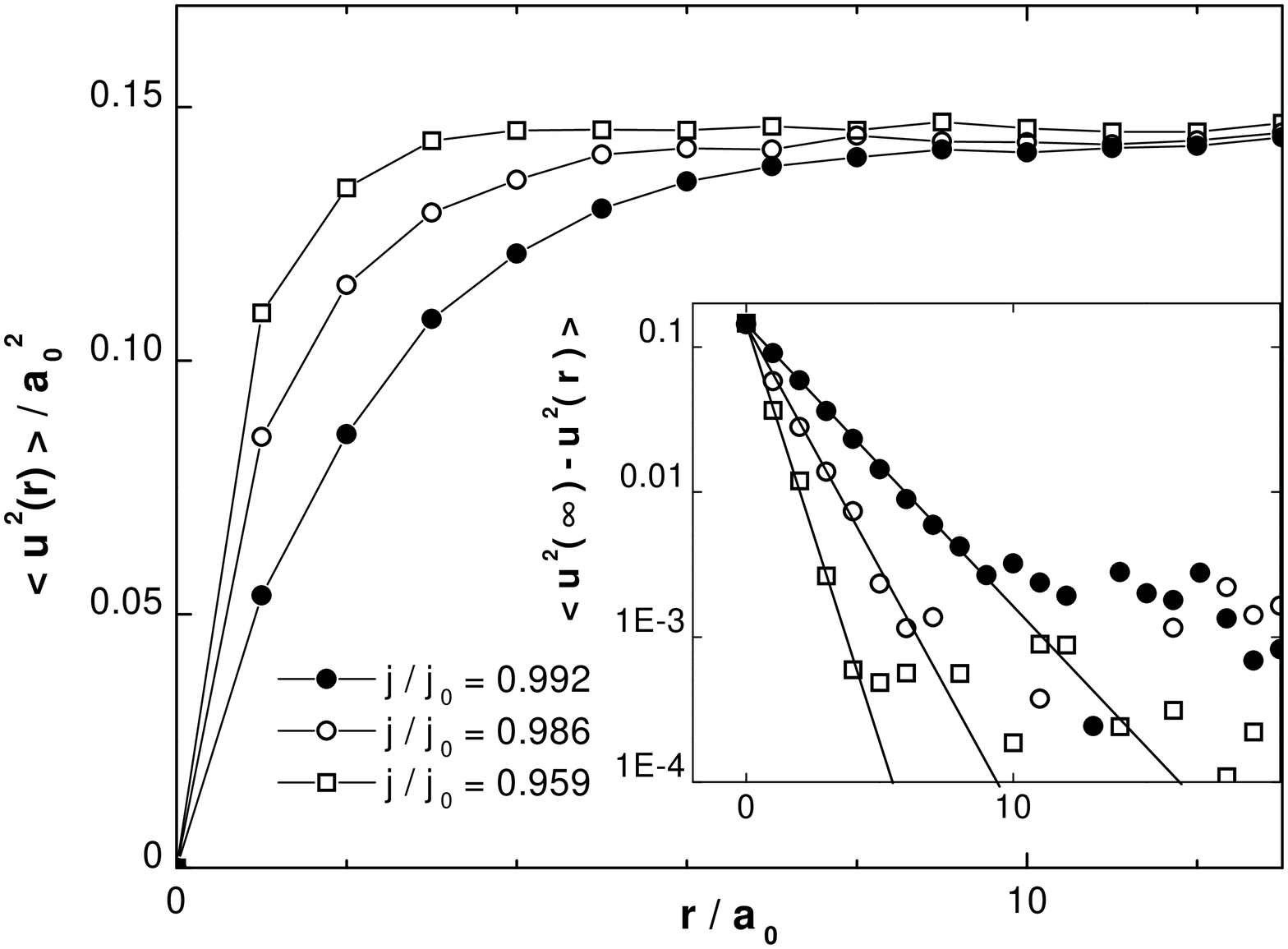}
} \caption{The correlation function for the displacement field.
($B = 3950$ Oe)} \label{u2}
\end{figure}

\begin{figure}
\resizebox{0.5\textwidth}{!}{%
  \includegraphics{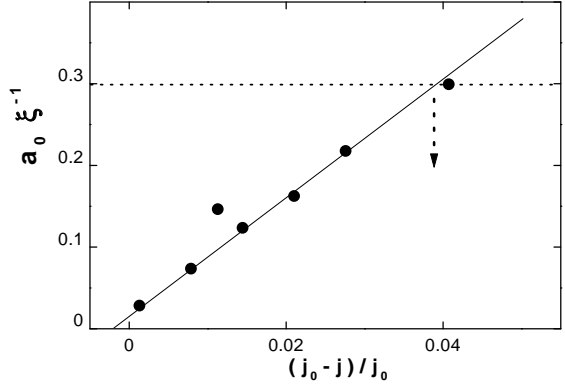}
} \caption{$B = 3950$ Oe. Correlation length for the positional
order, as obtained from the data in Fig.\ref{u2}. The full line
is a linear fit. The dotted line represents $<u^{ 2}> = c_{
L}^{2}\;a_0$ with $c_L=0.2$. The melting point, as indicated by
the dotted arrow, is shown in Fig.\ref{159} also.} \label{length}
\end{figure}

In order to characterize the continuous transition, let us
examine some autocorrelation functions which are classical tools
for the study of solids and liquids. The average hexatic order
parameter does not provide an accurate characterization of the
intermediate regime: as may be seen in Fig. \ref{159}, following
a sharp drop at $j=j_o$, there is no significative change at
lower $j$ where the density of defects however still exhibits
significant variations. The \textit{spatial correlations} of the
hexatic parameter carry more useful information\cite{nelson79}.
The correlator $<\Psi_{6}(0)\;\Psi_{6}^{*}(r)>_{r}$ for the data
in Fig.\ref{159} is shown in Fig.\ref{psi6}. Besides the
existence of a non zero background related to the non zero
averaged value $|<\Psi_{6}>|$, it reveals some additional short
range correlations of the orientational order, which extends to a
larger range as the system is closer to $j_o$. The oscillations
for small $r$ reflect the existence of a crystalline order within
this range: they are associated with the fluctuations of the
density autocorrelation function which come with the translational
symmetry breaking of the crystal order. This is confirmed by the
examination of the growth of the displacement field (actually a
positional correlation function): $<\mathbf{u}^{2}(r=n\; a_0)> =
<\sum_{j}
\mathbf{u}^{2}(n\;[\mathbf{r}_{j}-\mathbf{r}_{i}])>_{i}$ where $j$
denotes one of the nearest neighbors of vortex $i$ in a Delaunay
triangulation and $\mathbf{u}$ is the displacement field from the
periodic arrangement. As may be seen in Fig. \ref{u2}, there is an
exponential decay of the positional correlations, with a
diverging correlation length, $\xi(j)$, as one approaches $j_0$
from below (i.e. from larger 'temperatures'). Following ref.
\cite{nelson79}, one may then call this regime an 'hexatic liquid
crystal'. It is possible to track the positional correlation
length in Fig.\ref{u2} as one approaches $j_0$. The result is
diplayed in Fig.\ref{length} showing a divergence as $\xi =
\xi_0\;(1-j/j_0)^{-1}$, with the bare correlation length $\xi_0
\simeq 0.15\;a_0$. Recalling the Lindeman melting criterion,
$<u^{2}> = c_L^{2}\;a_0^{2}$ and the exponential increase of the
displacement field, $<u^{2}> = <u^{2}(\infty)> (1 -
\exp(-r/\xi))$, one may write an equivalent melting criterion for
the present case as $a_{0}/\xi =
\ln(1-c_{L}^{2}\;a_{0}^{2}/<u^{2}(\infty)>)$. The result obtained
using $c_{L}\approx 0.2$ and $<u^{2}(\infty)> \simeq 0.14$,
$a_{0}/\xi \approx0.3$, is displayed in Fig.\ref{159} and
\ref{length}. Although this quantitative result should be
considered with caution, due to the uncertainty on the effective
Lindeman number, this confirms that the solid has not melted in
the conventional way below the threshold value $j_0$.

\begin{figure}
\resizebox{0.5\textwidth}{!}{%
  \includegraphics{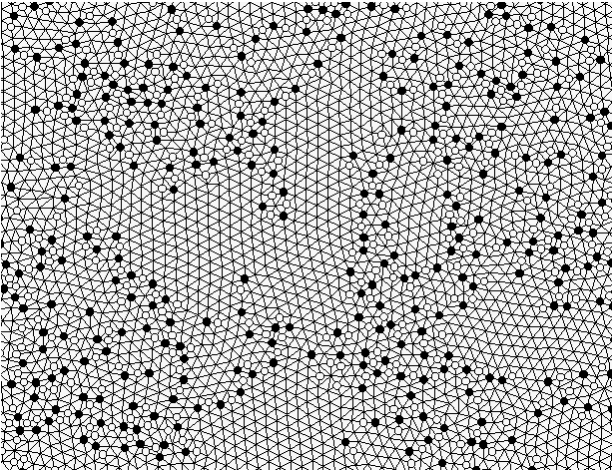}
} \caption{Positive - five-coordinated (white) and negative -
seven-coordinated (black) disclinations in a sample driven along
the vertical axis . ($B = 3950$ Oe, $j = 0.375$)} \label{chains}
\end{figure}

\begin{figure}
\resizebox{0.5\textwidth}{!}{%
  \includegraphics{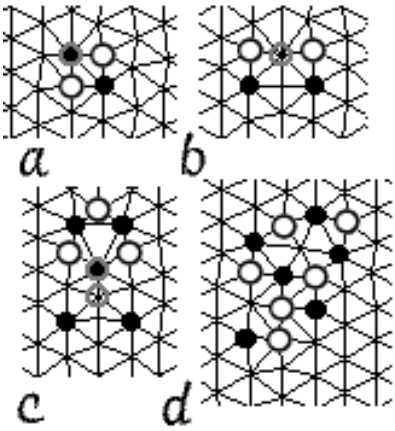}
} \caption{Early steps for the creation of defects ($B = 3950$
Oe, $j=0.375$. Nodes of the Delaunay triangulation lines are
vortices. Filled and opened circles are disclinations as in
Fig.\ref{chains}; gray circles indicate vortices which are
located in a potential minimum). Configurations a) and b) are
obtained consecutively as the result of the trapping of one
vortex. Configuration c) and d) are later steps. Vortices are
driven to the top of the figure.} \label{early}
\end{figure}

The observation that the transition is continuous at intermediate
induction and involves a proliferation of defects appeals for a
comparison with the KTHNY extension of the Kosterliz-Thouless
theory \cite{nelson79,young79}. The theory of dislocation mediated
melting of two-dimensional solids accounts for a continuous
transition, involving first unbinding of the dislocations leading
to the hexatic liquid, and then unbinding of the disclinations
leading to the regular liquid. Here, dislocations do not first
dissociate at $j_0$ to form an homogeneous 'plasma'. Rather, they
tend to form chains of alternating positive (five-coordinated)
and negative (seven-coordinated) disclinations which proliferate
in the liquid phase. As a result, unbounded dislocations and
disclinations remain marginal (Fig. \ref{chains}). Correlations
between dislocations were also reported in ref.\cite{ryu96} where
free dislocations, although not bound in chains, formed quenched
patterns moving with the average flux flow. In order to explain
the formation of these chains, the examination of the early
creation of defects in a driven crystal may be useful. Snapshots
of the earlier defects detected in a sample driven in the
intermediate region in Fig.\ref{159} are displayed in
Fig.\ref{early}. After a dislocation pair with opposite Burger
vectors has been created by the pinning of one vortex (a), it is
seen that the dislocations quickly arrange to form rings of
diameter $\sim 2\; a_0$ (c) and then larger loops (d).
Remarkably, the composite defects reflect the external force
anisotropy as soon as the dislocations dissociate (b): this
results from the plastic mechanism at work to create these
defects. This is also a direct evidence that the correlations in
the pseudo random force cannot be neglected in their formation.
The relation between these initial stages and the formation of
chains is not completely clear. A possible mechanism is the
stretching of elementary loops as in Fig.\ref{early}d, as the
vortices making disclinations appear to become more easily pinned
than the regular ones. This would make the long-range 'random
force' correlations a key ingredient in the chain formation
again. Equivalent rates for the growth and the annihilation of
the chains would then account for the existence of a stationary
regime intermediate between the crystal and the liquid.

In conclusion, it is found that a vortex lattice driven on dense
quasi-point pins shows a continuous transition between the
crystal and the liquid at intermediate induction, while first
order otherwise. The binding of the disclinations in chains is
proposed as a key mechanism to account for the existence of the
continuous transition.

Simulations have been performed on the cluster of the Centre de
Ressources Informatiques de l'Universite Paris-Sud (CRI).
%
%
%

\newpage
\newpage

\end{document}